\documentclass[proof]{WileyASNA-v1}

\articletype{Article Type}%

\received{26 April 2016}
\revised{6 June 2016}
\accepted{6 June 2016}

\begin{document}

\title{Isolated neutron stars as Science Validation for XMM2ATHENA\protect\thanks{XMM2ATHENA is a program funded by the European Union's Horizon 2020 research and innovation program.}: Ensuring robust data for future X-ray Astronomy}

\author[1,2]{Adriana Mancini Pires*}

\author[3]{Christian Motch}

\author[2]{Axel Schwope}

\author[2]{Iris Traulsen}

\author[4]{Jean Ballet}

\author[4]{Sudip Chakraborty}

\author[2]{David Homan}

\author[2]{Jan Kurpas}

\author[3]{Ada Nebot Gomez-Mor\`an}

\author[3]{Fran\c{c}ois-Xavier Pineau}

\author[5]{Hugo Tranin}

\author[5]{Natalie Webb}

\authormark{\textsc{Pires et al.}}

\address[1]{\orgdiv{Center for Lunar and Planetary Sciences, Institute of Geochemistry}, \orgname{Chinese Academy of Sciences}, \orgaddress{\state{99 West Lincheng Rd., 550051, Guiyang}, \country{China}}}

\address[2]{\orgdiv{Leibniz Institute for Astrophysics Potsdam}, \orgname{(AIP)}, \orgaddress{\state{Potsdam}, \country{Germany}}}

\address[3]{\orgdiv{Observatoire Astronomique de Strasbourg}, \orgname{CNRS, UMR 7550}, \orgaddress{\state{Strasbourg}, \country{France}}}

\address[4]{\orgdiv{Universit\'e Paris Saclay and Universit\'e Paris Cit\'e}, \orgname{CEA, CNRS, AIM}, \orgaddress{Gif-sur-Yvette, France}}

\address[5]{\orgdiv{IRAP, Universit\'e de Toulouse}, \orgname{CNRS, CNES}, \orgaddress{Toulouse, France}}

\corres{*Adriana Mancini Pires, 99 West Lincheng Rd., 550051, Guiyang, China. \email{adriana@mail.gyig.ac.cn}}

\abstract{
	The discovery of radio-quiet, X-ray thermally emitting isolated neutron stars (XINSs) in the ROSAT All-Sky Survey revealed a previously overlooked component of the neutron star population. Advancements in X-ray instrumentation and the availability of deep, wide-area optical surveys now enable us to explore XINSs at fainter X-ray fluxes and greater distances. In this study, we investigated candidates selected from the 4XMM-DR9 catalogue using XMM-Newton, focusing on long-term flux stability, spectral characterisation, and astrometry. By leveraging resources from the XMM2ATHENA project -- including updated catalogues, multiwavelength characterisation and machine learning classification -- we refined our understanding of this sample of soft X-ray emitters. Our findings enhance the characterisation of XINS candidates, laying the groundwork for more targeted investigations and future catalogue searches.
}

\keywords{stars: neutron, pulsars: general, X-rays: general, surveys, catalogs}

\jnlcitation{\cname{%
		\author{A.~M.~Pires}, 
		\author{C.~Motch}, and 
		\author{A.~Schwope}} (\cyear{2024}), 
	\ctitle{Isolated neutron stars as Science Validation for XMM2ATHENA: Ensuring robust data for future X-ray Astronomy}, \cjournal{Q.J.R. Meteorol. Soc.}, \cvol{2017;00:1--6}.}


\maketitle


\section{Introduction}\label{sec_intro}

A significant outcome of the ROSAT mission was the identification of a group of seven nearby (within a few hundred parsecs), middle-aged, radio-quiet isolated neutron stars \citep[][here dubbed XINSs]{2007Ap&SS.308..181H}. These sources are exceptional candidates for testing neutron star emission models due to their bright X-ray thermal radiation, independent distance estimates, and the absence of significant magnetospheric or accretion activity \citep{2020MNRAS.496.5052P}. Given that they are found in numbers comparable to those of young radio and gamma-ray pulsars within 1\,kpc, they may represent the only identified members of a larger, yet undetected, population of elusive neutron stars \citep{2008AIPC..983..331K,2008MNRAS.391.2009K}. Thus, discovering new candidates at faint X-ray fluxes is crucial for enhancing our understanding of their population characteristics and their relationships with other neutron star classes.

\begin{center}
	\begin{table*}[t]%
		\caption{\enspace Properties of XINS candidates from 4XMM-DR9\label{tab_dr9}}
		\centering
		\begin{tabular*}{500pt}{@{\extracolsep\fill}lrrcrrrrr@{\extracolsep\fill}}
			\toprule
			\textbf{Target} & \multicolumn{1}{c}{\textbf{RA}} & \multicolumn{1}{c}{\textbf{Dec}} & \textbf{Error} & \textbf{Flux$^{(1)}$} & \textbf{$m_{r}$}$^{(2)}$ & \multicolumn{3}{c}{\textbf{Hardness Ratios}$^{(3)}$} \\
			\cmidrule{7-9}
			4XMM & \multicolumn{1}{c}{(deg)} & \multicolumn{1}{c}{(deg)} & ($''$) & & & \textbf{HR$_1$} & \textbf{HR$_2$} & \textbf{HR$_3$}\\
			\midrule
			J123337.8$+$374127 & $188.40773$ & $+37.69089$ & $1.5$ & $2.9(4)$   & $>24.19$ & $0.71(10)$ & $-0.31(11)$ & $-0.86(20)$\\
			J140340.4$-$603007 & $210.91874$ & $-60.50209$ & $0.7$ & $1.98(12)$ & $>23.04$ & $0.85(3)$ & $-0.61(4)$ & $-0.99(7)$\\
			J010331.0$+$250540 & $15.87951$  & $+25.09457$ & $0.9$ & $1.93(20)$ & $>24.31$ & $0.19(8)$ & $-0.38(8)$ & $-0.99(6)$ \\
			J022141.5$-$735632 & $35.42309$  & $-73.94222$ & $0.9$ & $1.59(11)$ & $>25.01$ & $-0.663(21)$ & $-0.91(4)$ & $-0.95(24)$ \\			
			J225139.6$-$162748 & $342.91527$ & $-16.46335$ & $1.1$ & $1.10(17)$ & $>24.28$ & $0.15(11)$ & $-1.00(8)$ & $1\pm4$ \\
			\bottomrule
		\end{tabular*}
		\begin{tablenotes}
			\item Errors (in brackets) represent $1\sigma$ confidence levels. Targets are sorted by decreasing flux (\texttt{SC\_EP\_2\_FLUX}). Superscripts indicate: $^{(1)}$ catalogued EPIC flux in units of $10^{-14}$ erg s$^{-1}$ cm$^{-2}$ for the 0.5--1 keV band; $^{(2)}$ optical magnitude $5\sigma$ upper limit from the deepest available survey; $^{(3)}$ hardness ratios (HR), defined as the ratio of the difference to total counts across four of the five XMM-Newton energy bands: 0.2--0.5 keV, 0.5--1 keV, 1--2 keV, and 2--4.5 keV, with signals typically undefined in HR$_4$.
		\end{tablenotes}
	\end{table*}
\end{center}

In preparation for searches at the full sensitivity of the eROSITA All-Sky Survey \citep{2024A&A...682A..34M}, exploring serendipitous data from the XMM-Newton Observatory offers an excellent opportunity to test search algorithms and discover new XINSs beyond the Solar vicinity. Building on our previous experience in cross-correlating earlier releases of the XMM-Newton X-ray source catalogue \citep{2009A&A...504..185P,2017ASPC..512..165M}, we conducted a search for new XINS candidates in the fourth generation of the XMM-Newton catalogue, 4XMM-DR9 (\citealt{2022A&A...666A.148P}; see also recent XINS searches in \citealt{2022MNRAS.509.1217R,2024ApJ...961...36D,2024A&A...687A.251K}). Over the past years, the brightest unknown sources surviving our selection procedure were prioritised for follow-up XMM-Newton observations. In \cite{2022A&A...666A.148P}, we reported the discovery of a new cooling XINS identified through this programme.

The work presented here validates the tools and catalogues developed within the XMM2ATHENA project \citep{2023AN....34420102W}. In synergy with the XMM-Newton Survey Science Centre \citep{2001A&A...365L..51W}, XMM2ATHENA enhances the scientific value of XMM-Newton data by enabling improved source detection methods, real-time X-ray transient monitoring, multiwavelength counterpart identification, and machine learning-based source classification \citep[e.g.][]{2022A&A...657A.138T,2024A&A...687A.250Q,2024A&A...683A.172M}. 
In this paper, we examine the sample of XINS candidates from 4XMM-DR9 followed up with second-epoch XMM-Newton observations (Section~\ref{sec_followup}). Section~\ref{sec_x2a} discusses the validation of XMM2ATHENA tools, and Section~\ref{sec_summary} outlines future research directions. A more detailed investigation of the 4XMM-DR9 candidate sample and the properties of the XINS population will be presented in a forthcoming publication.

\begin{center}
	\begin{table*}[t]%
		\centering
		\caption{\enspace Summary of XMM-Newton observations and astrometry\label{tab_obs}}%
		\tabcolsep=0pt%
		\begin{tabular*}{500pt}{@{\extracolsep\fill}lccccrcrrrcr@{\extracolsep\fill}}
			\toprule
			\textbf{INSC} & \textbf{Obs.~Date} & \textbf{Mode}\tnote{$^{(1)}$} & \textbf{Filter}\tnote{$^{(2)}$} & \textbf{$\theta$}\tnote{$^{(3)}$} & \multicolumn{2}{c}{\textbf{GTI}\tnote{$^{(4)}$}} & \multicolumn{1}{c}{\textbf{Counts}}\tnote{$^{(5)}$} & \multicolumn{3}{c}{\textbf{Boresight correction}\tnote{$^{(6)}$}} & \multicolumn{1}{c}{\textbf{$r$}\tnote{$^{(7)}$}}\\
			\cmidrule{6-7}\cmidrule{9-11}
			& & & & \multicolumn{1}{c}{($'$)} & \multicolumn{1}{c}{(s)} & \multicolumn{1}{c}{(\%)} & & \multicolumn{1}{c}{($\Delta\alpha,''$)} & \multicolumn{1}{c}{($\Delta\delta,''$)} & $N_{\rm ref}$ & \multicolumn{1}{c}{($''$)} \\
			\midrule
			J1233 & 2003-11-24 & FF & Thin & $16$  & $11\ 667$ & 94 & $104(13)$ & $0.45(40)$ & $-1.22(40)$ & 69 & 0.4\\
			& 2022-11-20 & FF & Thin & $0.9$ & $30\ 267$ & 95 & $1820(50)$ & $+0.24(22)$ & $+0.56(23)$ & 101 & 5.8\\
			J1403 & 2003-07-19 & FF & Thick & $7.3$ & $47\ 200$ & 72 & $980(40)$ & $-0.46(35)$ & $+0.51(34)$ & 107 & 2.7\\
			& 2024-01-18 & LW/FF & Medium & $0.8$ & $36\ 400$ & 82 & $910(40)$ & $+0.90(40)$ & $<0.40$ & 52 & 0.5\\
			J0103 & 2016-07-02 & FF & Thin & $11$  & $15\ 467$ & 77 & $205(18)$ & $-0.68(50)$ & $<0.40$ & 67 & 0.7\\
			& 2023-01-24 & FF & Thin & $0.8$ & $8\ 700$  & 29 & $83(14)$ & $-2.30(60)$ & $+1.00(60)$ & 27 & 1.1\\
			J0221 & 2012-02-09 & FF & Medium & $9.5$ & $28\ 867$ & 90 & $1310(40)$ & $-0.46(28)$ & $-0.51(26)$ & 60 & 1.7\\
			& 2021-07-09 & FF & Thin   & $0.9$ & $29\ 533$ & 71 & $2110(50)$ & $<0.30$ & $-1.33(29)$ & 67 & 1.9\\
			J2251 & 2015-11-27 & FF & Medium & $9.7$ & $11\ 167$ & 94 & $115(14)$ & $<0.60$ & $<0.60$ & 50 & 1.4\\
			& 2022-05-15 & FF & Med./Thin & $0.8$ & $19\ 267$ & 61 & $230(20)$ & $+0.70(30)$ & $-0.75(25)$ & 72 & 1.4\\			
			\bottomrule
		\end{tabular*}
		\begin{tablenotes}
			\item Errors (in brackets) indicate $1\sigma$ confidence levels. Superscripts denote: $^{(1)}$ EPIC science mode (full frame, FF, or large window, LW); $^{(2)}$ optical blocking filters; $^{(3)}$ off-axis angle in arcminutes; $^{(4)}$ good-time intervals in seconds and as a percentage of the total observation duration; $^{(5)}$ total source counts in the 0.2--2\,keV energy band; $^{(6)}$ boresight corrections in right ascension ($\Delta\alpha$) and declination ($\Delta\delta$) in arcseconds; $^{(7)}$ angular distance from the corrected position to the 4XMM-DR9 coordinates.
		\end{tablenotes}
	\end{table*}
\end{center}

\section{XMM-Newton follow-up}\label{sec_followup}

As of this writing, five selected targets have been observed with XMM-Newton. Table~\ref{tab_dr9} presents the overall properties of these sources as listed in the 4XMM-DR9 catalogue \citep{2020A&A...641A.136W}, including their positions, positional errors, fluxes, and hardness ratios. Additionally, we provide $5\sigma$ optical upper limits in the $r$ band based on the deepest available survey for each candidate field. 

Table~\ref{tab_obs} summarises the five observation pairs, organised by XINS candidate (INSC; identified by the first four digits of their IAU designations, JXXXX). This table includes the observation date, science mode, and EPIC filter used. The good-time intervals (GTI), shown in seconds and as a percentage of the total observation duration, are filtered for high background activity and averaged across all active EPIC cameras. We also report the total source counts in the 0.2--2 keV energy band and the off-axis angle $\theta$ (in arcminutes) for each source.

\begin{figure*}[t]
	\centerline{
		\includegraphics[width=0.25\textwidth]{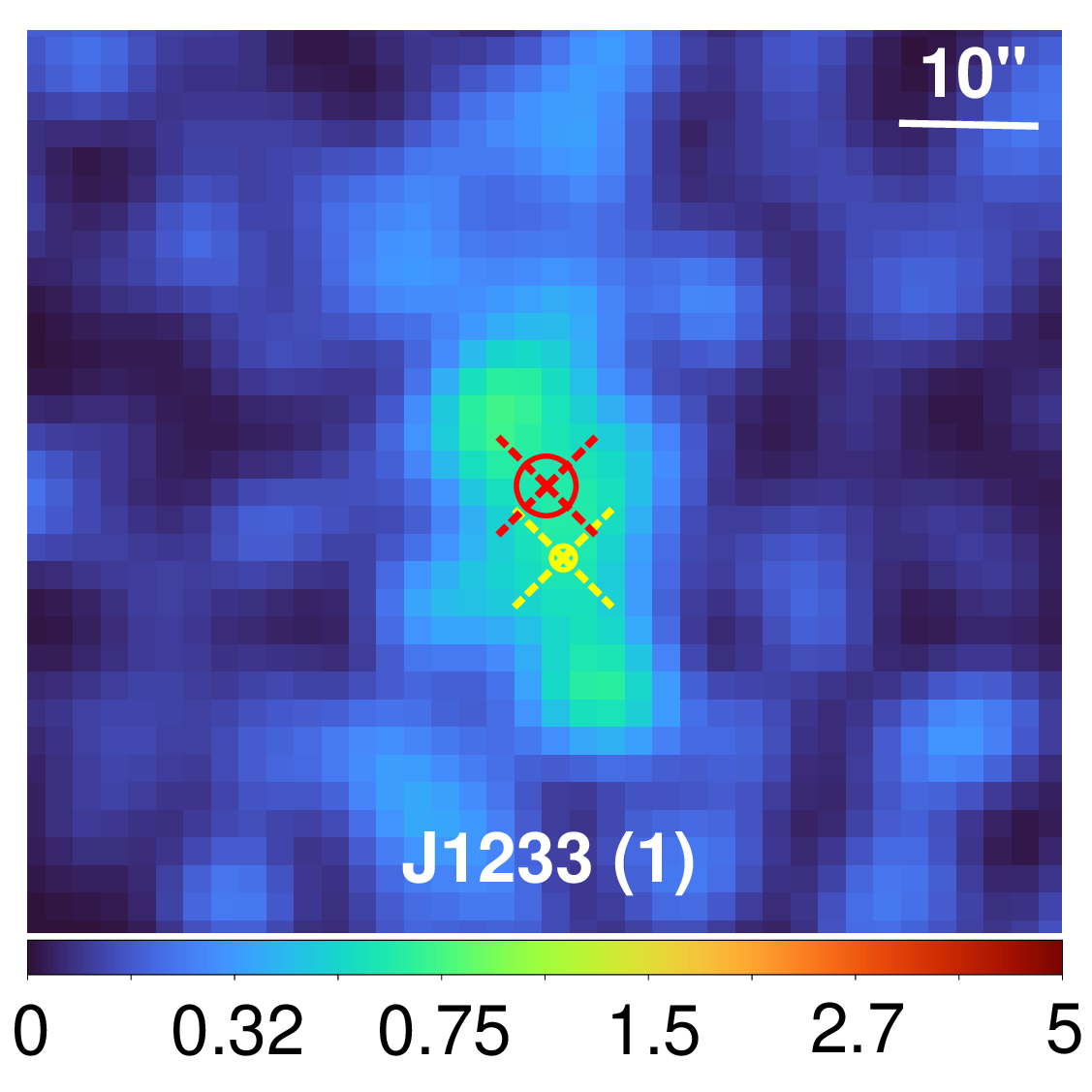}
		\includegraphics[width=0.25\textwidth]{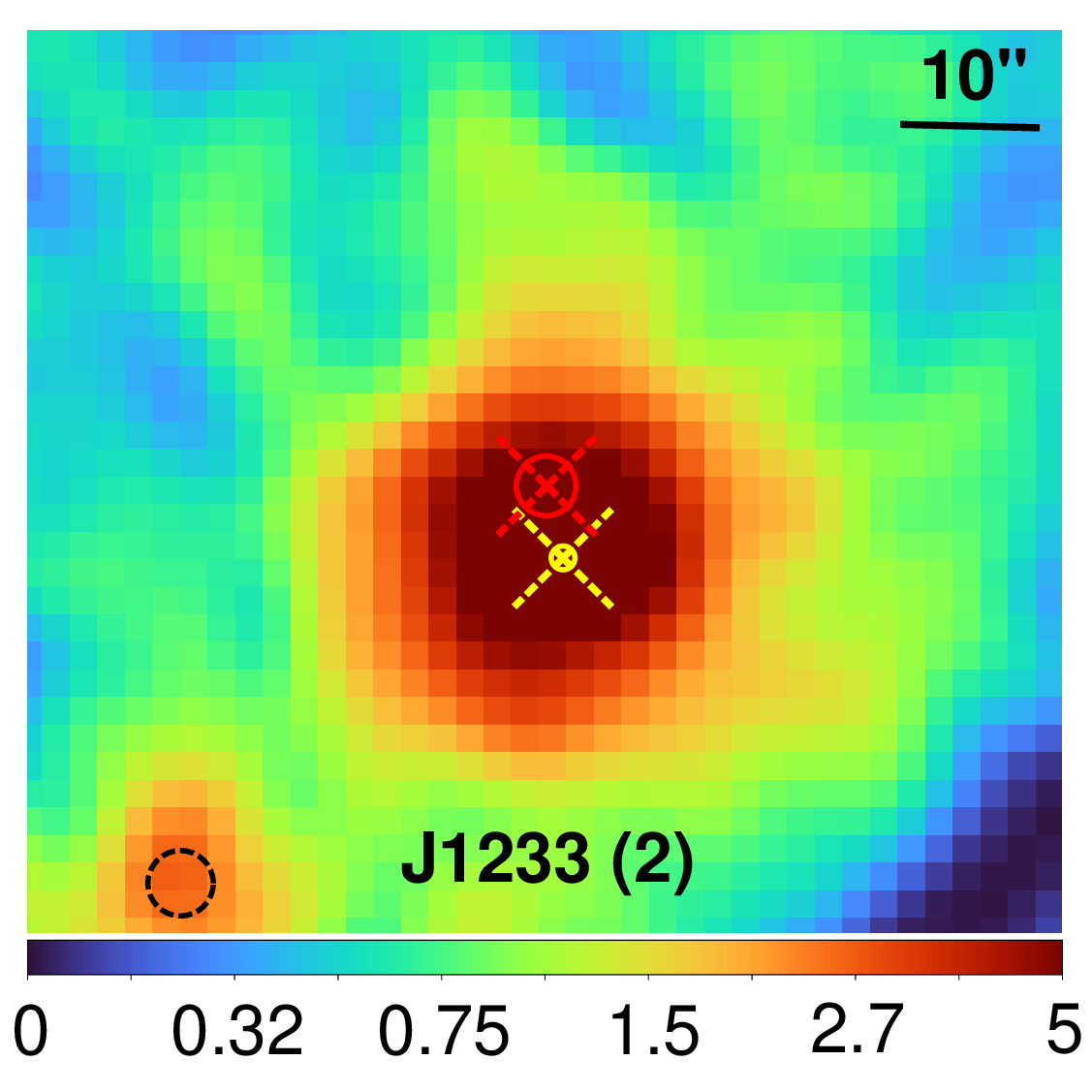}
		\includegraphics[width=0.25\textwidth]{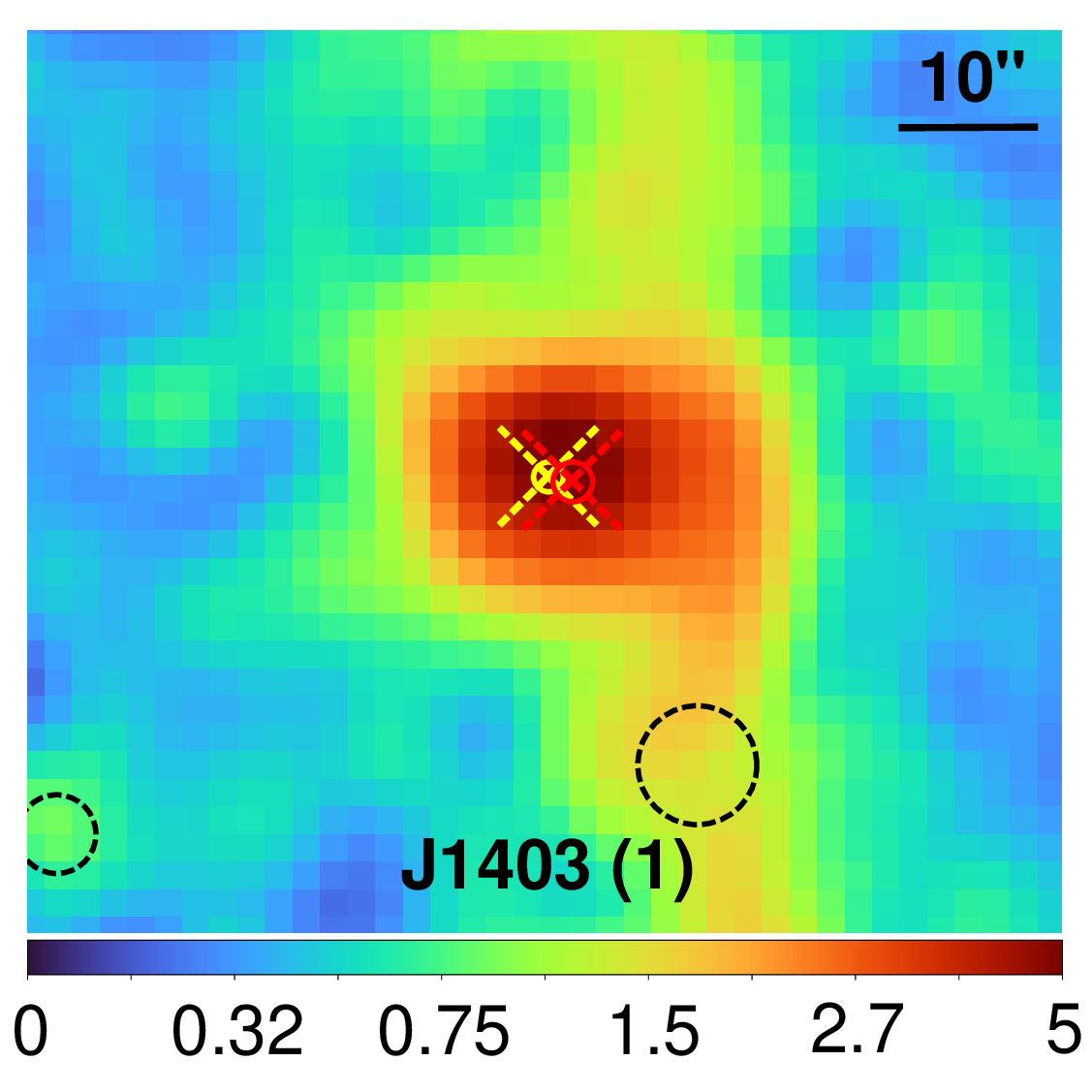}
		\includegraphics[width=0.25\textwidth]{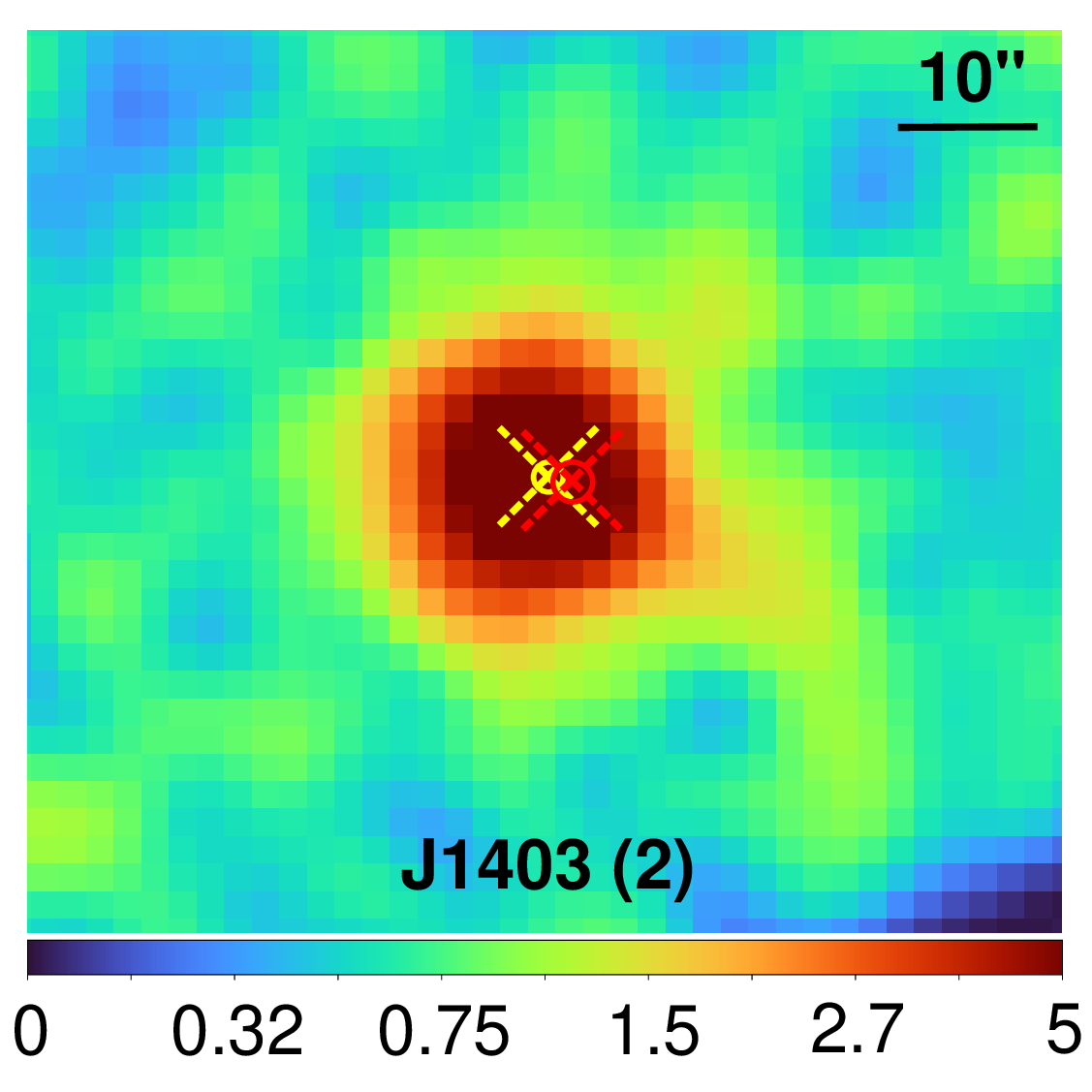}}
	\caption{\enspace First- and second-epoch X-ray images (EPIC pn, 0.2--12 keV) of the fields of INSC J1233 (left) and J1403 (right). Astrometrically corrected target positions are indicated by 90\% confidence level error circles with crosses: red for serendipitous observations and blue/green for aimpoint observations. Black dashed outlines are X-ray sources detected within the field of view.\label{fig_shifted}}
\end{figure*}

We conducted source detection on both individual and stacked observation pairs using \texttt{edetect\_stack}, which enabled us to obtain more accurate source parameters through stacked PSF fitting \citep{2020A&A...641A.137T}. To refine the astrometry, we applied \texttt{eposcorr} to cross-correlate the EPIC X-ray source positions with those from the optical Guide Star Catalogue~2.4.2 \citep{2008AJ....136..735L}, within a 15$'$ radius of the nominal pointing coordinates. Table~\ref{tab_obs} shows the small positional offsets in right ascension ($\Delta\alpha$) and declination ($\Delta\delta$), in arcseconds, derived from boresight corrections using $N_{\rm ref}$ X-ray and optical matches. These offsets were applied to the attitude files of individual observations, aligning the events of each epoch prior to stacked detection. 

Second-epoch observations improved source localisation by factors of 2 to 4. For most XINS candidates, there was no significant shift in their astrometrically corrected positions relative to the 4XMM-DR9 coordinates (see $r$ in Table~\ref{tab_obs}). Exceptions include J1233, which was detected only by the EPIC pn camera at a large off-axis angle (Fig.~\ref{fig_shifted}, first panel), and J1403, which was affected by an out-of-time (OOT) event (Fig.~\ref{fig_shifted}, third panel). The updated position of J1233, shifted by $5.8''$, strongly suggests an association between the X-ray source and the radio galaxy NVSS~J123337+374122 \citep{2020ApJS..247...53K}. Despite a shift of $2.7''$, J1403 remains without identified counterparts, retaining its status as an XINS candidate (see also Section~\ref{sec_x2a}).

\begin{center}
	\begin{table*}[t]%
		\caption{\enspace Results of spectral analysis\label{tab_spec}}
		\centering
		\begin{tabular*}{520pt}{@{\extracolsep\fill}lrrcrrrrrrc@{\extracolsep\fill}}
			\toprule
			\textbf{INSC} & \multicolumn{3}{c}{\textbf{(I) Constant emission (power-law)}} & \multicolumn{7}{c}{\textbf{(II) Independent epochs (blackbody)}} \\
			\cmidrule{2-4}\cmidrule{5-11}
			& \multicolumn{1}{c}{$N_{\rm H,21}^{\rm PL\dagger}$} & \multicolumn{1}{c}{$\Gamma$} & \multicolumn{1}{c}{$f_{\rm X,14}^{\rm PL\ddag}$} & \multicolumn{1}{c}{$N_{\rm H,21}^{\rm BB\dagger}$} & \multicolumn{1}{c}{$kT^{(1)}$} & \multicolumn{1}{c}{$kT^{(2)}$} & \multicolumn{1}{c}{$R_{1\textrm{kpc}}^{\ast~(1)}$} & \multicolumn{1}{c}{$R_{1\textrm{kpc}}^{\ast~(2)}$} & \multicolumn{1}{c}{$f^{\rm BB\ddag~(1)}_{\rm X,14}$} & \multicolumn{1}{c}{$f^{\rm BB\ddag~(2)}_{\rm X,14}$} \\
			& & & & & \multicolumn{1}{c}{(eV)} & \multicolumn{1}{c}{(eV)} & \multicolumn{1}{c}{(km)} & \multicolumn{1}{c}{(km)} & & \\
			\midrule
			J1233 & $11.2(9)$ & $7.7(5)$ & $5.45_{-0.18}^{+0.16}$ & $3.6(5)$ & $180_{-16}^{+18}$ & $164(8)$ & $4^\star$ & $0.56_{-0.10}^{+0.13}$ & $4.9(6)$ & $5.43_{-0.18}^{+0.15}$ \\
			J1403 & $17.0_{-1.3}^{+1.4}$ & $12.6(8)$ & $2.61_{-0.06}^{+0.09}$ & $6.8(7)$ & $94(5)$ & $99_{-5}^{+6}$ & $8^\star$ & $3.4_{-0.9}^{+1.4}$ & $2.5(4)$ & $2.30_{-0.09}^{+0.10}$ \\
			J0103 & $1.4_{-0.6}^{+0.8}$ & $3.5_{-0.5}^{+0.6}$ & $3.7_{-0.4}^{+0.9}$ & $0.05^\star$ & $178_{-11}^{+12}$ & $260_{-40}^{+50}$ & $2.6^\star$ & $0.059_{-0.018}^{+0.025}$ & $4.3(4)$ & $1.50(25)$ \\
			J0221 & $1.77(11)$ & $8.47_{-0.25}^{+0.26}$ & $11.44_{-0.24}^{+0.29}$ & $0.70_{-0.08}^{+0.09}$ & $58.7_{-1.9}^{+2.0}$ & $61.5(1.9)$ & $10^\star$ & $7.9_{-1.1}^{+1.3}$ & $12.4_{-0.5}^{+0.4}$ & $11.6_{-0.3}^{+0.4}$ \\
			J2251 & $0.8_{-0.3}^{+0.4}$ & $4.4_{-0.4}^{+0.5}$ & $2.56_{-0.3}^{+0.28}$ & $<0.06$ & $123_{-9}^{+10}$ & $119(7)$ & $3.2^\star$ & $0.37_{-0.05}^{+0.06}$ & $3.3(4)$ & $2.28(18)$ \\			
			\bottomrule
		\end{tabular*}
		\begin{tablenotes}
			\item Errors (in brackets) indicate $1\sigma$ confidence levels. Superscripts denote: $^\dagger$ column density in units of $10^{21}$\,cm$^{-2}$; $^\ddag$ observed source flux in units of $10^{-14}$\,erg\,s\,cm$^{-2}$ in the 0.2--12\,keV energy band; $^\ast$ size of the blackbody emission region normalised to a distance of 1\,kpc; $^\star$ parameter is unconstrained; $^{(1)}$ values in the first epoch; $^{(2)}$ values in the second epoch.
		\end{tablenotes}
	\end{table*}
\end{center}

For the spectral analysis, we performed simultaneous fits of the EPIC spectra for each X-ray source using XSPEC~12.13.1. A renormalisation factor was included to account for cross-calibration uncertainties between instruments, and the \texttt{tbabs} absorption model with cross-sections from \citet{2000ApJ...542..914W} was applied. First, we fitted an absorbed power-law model, assuming the parameters remained constant between epochs by tying them across datasets. This approach facilitates comparison with results from source detection in \textit{spectral} mode, a new feature of XMM2ATHENA (Section~\ref{sec_x2a}). In a separate analysis, we applied a blackbody model, which better represents the soft X-ray emission of XINSs. To explore potential long-term variability, we allowed the temperature and normalisation of the blackbody model to vary independently between epochs, while keeping the column density fixed.

The results are summarised in Table~\ref{tab_spec}. For each method (labelled I and II), we present the best-fit spectral parameters: column density ($N_{\rm H}$), photon index ($\Gamma$) for the power-law model, blackbody temperature and radius ($kT$ and $R$), and the corresponding fluxes. As expected, the photon indices are steep, indicating an intrinsically thermal continuum. The column density values from model I exceed the line-of-sight values in all cases \citep{2016A&A...594A.116H}. The fit quality in both approaches is generally good, with null-hypothesis probabilities typically exceeding 50\%. A notable exception is INSC J1233, which displays an energy distribution consistent with a hot and thin thermal plasma, confirming the source as the X-ray counterpart of the active radio galaxy.

\begin{figure*}[t]
	\centerline{
		\includegraphics[width=\textwidth]{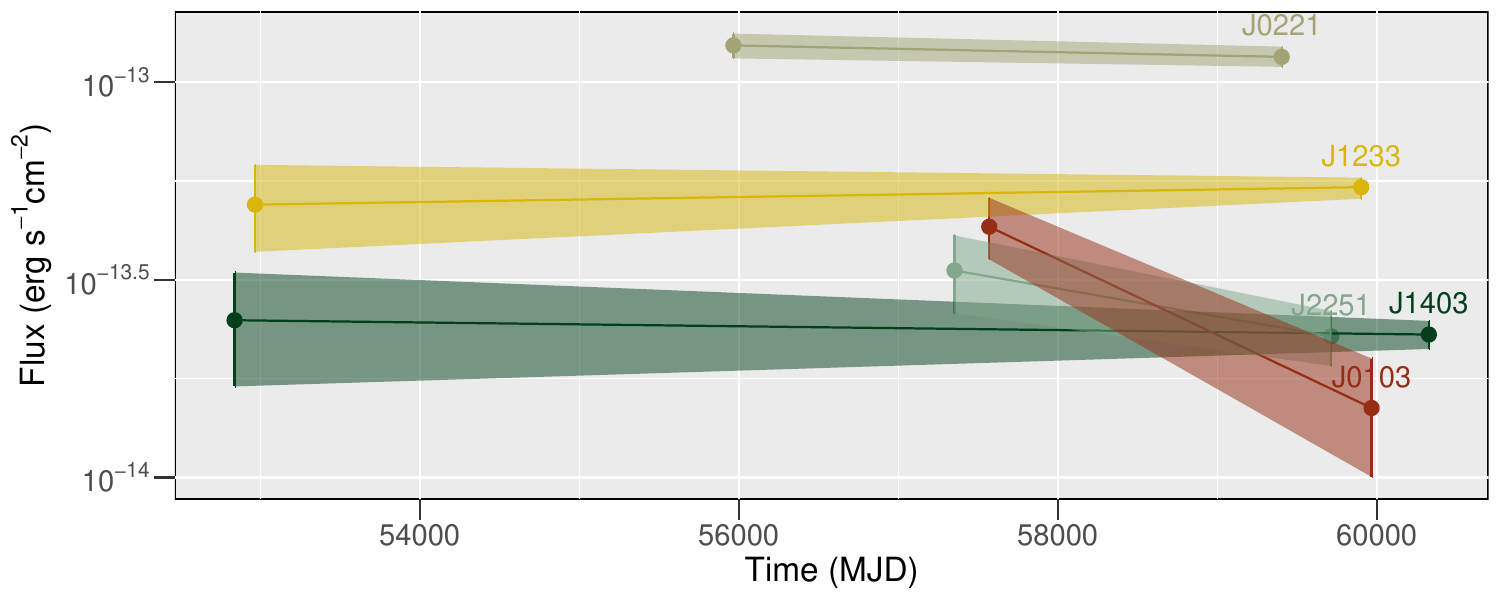}}
	\caption{\enspace Flux evolution of the five XINS candidates with $2\sigma$ confidence levels, highlighting the variability observed in J0103.\label{fig_fxevol}}
\end{figure*}

In all cases, approach II shows no significant spectral variability within errors, with blackbody parameters remaining consistent within $2\sigma$. The observed fluxes listed in Table~\ref{tab_spec} are primarily derived using \texttt{cflux} from the best-fit models. When there is significant flux dispersion across the three EPIC cameras within an epoch -- such as in the pn exposure of J1403 affected by the OOT event -- the median value is used instead. For all sources except J0103, these fluxes remain stable across epochs (Fig.~\ref{fig_fxevol}). The analysis shows that J0103 is now nearly three times dimmer than in the first epoch, suggesting a possible tidal-disruption event and arguing against an XINS classification. However, it is worth noting that the follow-up observation of J0103 suffered over 70\% data loss due to high background levels (see Table~\ref{tab_obs}).

\section{XMM2ATHENA validation}\label{sec_x2a}

Source detection in the analysis of XMM-Newton observations is performed through simultaneous PSF fitting across multiple energy bands, cameras, and overlapping sky areas. The method developed by XMM2ATHENA for the Enhanced Stacked Catalogue (ESC) simplifies the detection process by assuming that the X-ray emission follows a basic spectral model and remains constant over time. This approach reduces the degrees of freedom in fitting, leading to better handling of detection thresholds. It also enhances sensitivity to faint sources and directly provides spectral information for each catalogue entry.

In the enhanced approach (\textit{spectral}, as opposed to rates fitting), energy conversion factors (ECFs) between count rates and observed fluxes play a much more central role than in previous XMM-Newton catalogue production. The fitting routines maximise the likelihood calculation of the source's PSF by considering its spectrum, detector position, energy, and response. This is made possible through newly provided pre-computed grids of ECF and per-band flux ratio values, a new code module for interpolating grid values, and modifications to the fitting routines that incorporate parameter coupling and reduce the degrees of freedom necessary for spectral maximum likelihood calculations.

\begin{center}
	\begin{table*}[t]%
		\caption{\enspace Results of XMM2ATHENA validation\label{tab_x2a}}
		\centering
		\begin{tabular*}{500pt}{@{\extracolsep\fill}lccccc@{\extracolsep\fill}}
			\toprule
			\textbf{Parameter} & \textbf{J1233} & \textbf{J1403} & \textbf{J0103} & \textbf{J0221} & \textbf{J2251}\\
			\midrule
			Detection likelihood & 3340 & 1870 & 420 & 11570 & 370\\
			Extent (arcsec) & $3.24(19)$ & 0 & 0 & 0 & 0\\
			Flux ($10^{-14}$\,erg\,s$^{-1}$\,cm$^{-2}$) & $8.42(13)$ & $2.72(8)$ & $2.55(13)$ & $19.70(29)$ & $1.95(11)$\\
			$N_{\rm H}$ ($10^{21}$\,cm$^{-2}$) & $5.04(14)$ & $4.44(15)$ & $2.03(18)$ & $0.171(9)$ & $1.34(15)$\\
			Photon index & $5^\star$ & $5^\star$ & $3.62(12)$ & $5^\star$ & $5^\star$\\			
			Non-matching probability (\%) & $<1$ & $93$  & $100$ & $100$ & $2$\\
			X-ray-to-optical flux ratio & $23$ & $>10$ & $>40$ & $>300$ & $4$\\
			\midrule
			Classification or prediction & Radio galaxy/AGN  & XINS cand. & TDE cand./Quasar & Confirmed XINS & Quasar\\
			\bottomrule
		\end{tabular*}
		\begin{tablenotes}
			\item Errors (in brackets) indicate $1\sigma$ confidence levels. $^{\star}$ Parameter is unconstrained.
		\end{tablenotes}
	\end{table*}
\end{center}

We applied the new software to stacked observations from the XINS programme, with results summarised in Table~\ref{tab_x2a}. Notably, INSC J1233, a known contaminant in the XINS candidate sample, was correctly classified as an extended source, distinguishing it from the other candidates. The spectral parameters derived from the new method align closely with those from spectral analysis, highlighting its reliability. However, the ECF grid is currently limited to photon indices between 0 and 5, and most candidates -- except for the long-term variable J0103 -- exceed this range, restricting full optimisation of their spectral parameters. Nevertheless, for these very soft sources, the new method provides a more accurate approximation of the true flux compared to the standard detection approach, which assumes a power-law model with $\Gamma=1.7$ and $N_{\rm H}=3\times10^{20}$\,cm$^{-2}$ when converting counts to fluxes \citep{2009A&A...496..879M}.

We cross-correlated the updated source positions with optical and near-infrared objects from several catalogues, including Gaia DR3 \citep{2021A&A...649A...1G}, Legacy Survey DR10 \citep{2019AJ....157..168D}, unWISE \citep{2019ApJS..240...30S}, and DeCAPS2 \citep{2023ApJS..264...28S}, using the ARCHES tool\footnote{\texttt{http://serendib.unistra.fr/ARCHESWebService}}. Matching probabilities were calculated based on angular distance, local object distribution, and positional accuracy, following the method outlined by \cite{2017A&A...597A..89P} and refined in XMM2ATHENA. Table~\ref{tab_x2a} presents the resulting multi-catalogue probabilities. The identification of highly probable counterparts for J1233 and J2251 effectively rules out an XINS nature for these sources (noting that Legacy Survey DR10 was not publicly available during the initial search for XINS candidates). Additionally, cross-matching with updated X-ray coordinates revealed a likely near-ultraviolet counterpart for J1233, GALEX J123338.0+374124, with a matching probability of 93\% and a magnitude of 22.6(4).

Finally, we applied an enhanced version of the machine learning classifier described in \citet{2022A&A...657A.138T} to the XINS sample. Consistent with the \textit{spectral} source detection, variability assessment, and ARCHES characterisation, the candidates J0221 and J1403 are the most likely XINSs, while the other sources are classified as extragalactic based on the spatial, spectral, and timing properties of both the X-ray sources and their multiwavelength counterparts (Table~\ref{tab_x2a}).

\begin{figure*}[t]
	\centerline{
		\includegraphics[width=\textwidth]{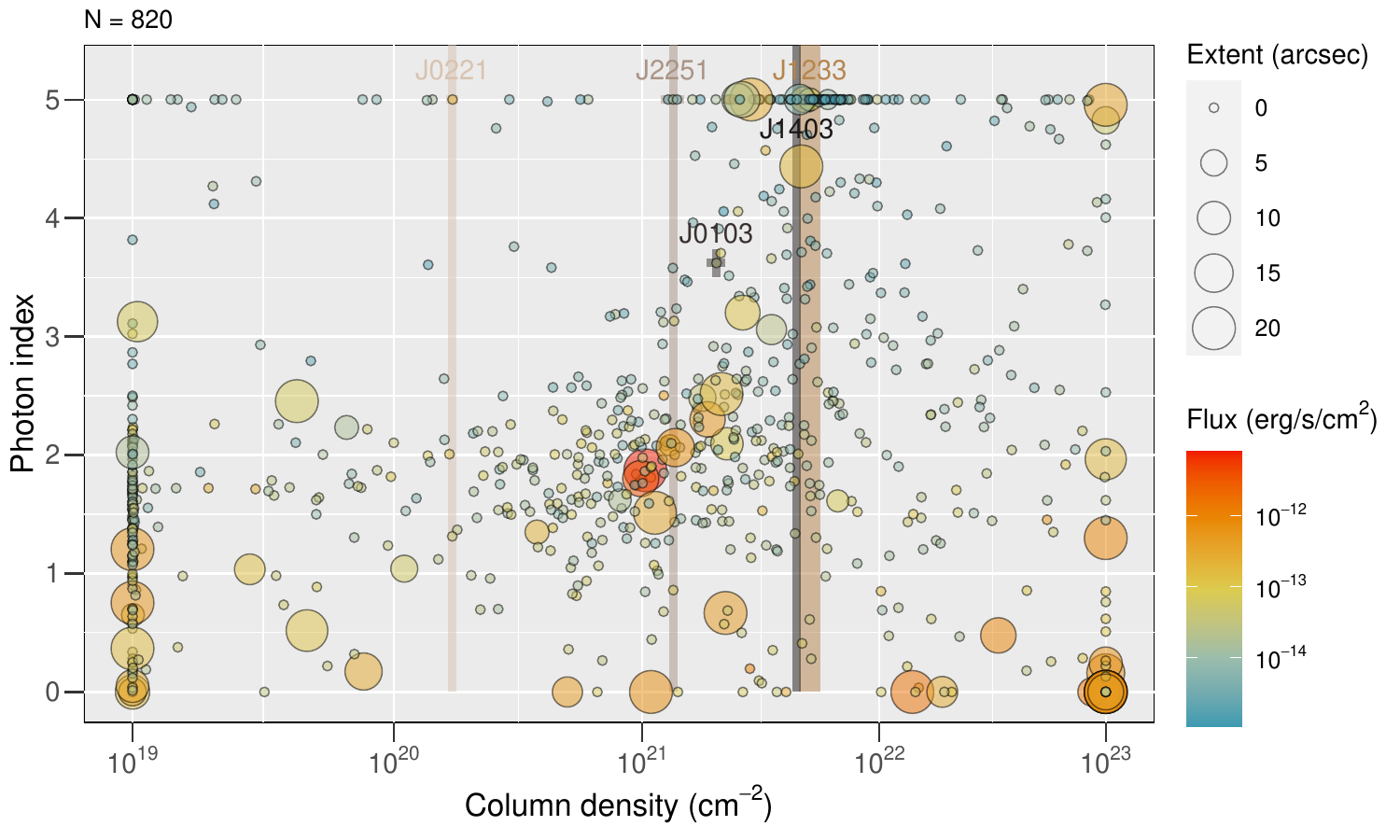}}
	\caption{\enspace Photon index ($\Gamma$) as a function of column density ($N_{\rm H}$) for 820 ESC-detected X-ray sources across five sky fields. Data points with labels and error bars (with unconstrained $\Gamma$ for most sources) highlight the five XINS candidates.\label{fig_nhgamma}}
\end{figure*}

In Fig.~\ref{fig_nhgamma}, we present a scatter plot of the ESC spectral parameters, specifically $\Gamma$ versus $N_{\rm H}$, for the 820 sources detected across the five stacked sky fields in which the targets are located. The size and colour dimensions reflect the sources' extent and flux, respectively. Overall, the analysis demonstrates a more efficient approach to selecting soft sources compared to traditional reliance on hardness ratios. For future searches, XINS candidates could be identified using criteria such as $\Gamma=5$ and \texttt{EXTENT=0}, combined with ARCHES non-association probabilities exceeding a set threshold.

\section{Discussion and outlook}\label{sec_summary}

In this work, we employed the newly developed XMM2ATHENA software to validate source detection methods for stacked X-ray observations from the 4XMM-DR9 XINS programme. Our focus extended beyond source detection; we also validated multiwavelength characterisation and machine learning classification tools that supported and complemented our findings.

A straightforward selection of extremely soft point sources without optical or near-infrared counterparts was sufficient to identify all viable XINS candidates in the programme. The consistency between spectral parameters obtained from source detection and those from spectral analysis confirms the reliability of this new method.

The advantages of re-observing candidates at the aimpoint and stacking are evident, demonstrated by improved localisation and the correct classification of the contaminant INSC J1233 as an extended source. The identification of a faint optical counterpart in Legacy Survey DR10 for J2251, along with long-term flux variability in J0103, suggests that these sources are unlikely to be XINSs. Additionally, candidate J1403 requires deeper optical limits and further follow-up to confirm its status as a potential XINS.

In recent years, there has been renewed interest in identifying XINSs that conventional radio and gamma-ray pulsar surveys have overlooked. Initial results from flux-limited searches using data from the Western Galactic hemisphere of the eROSITA All-Sky Survey are beginning to emerge \citep{2024A&A...687A.251K,2023A&A...674A.155K,2024A&A...683A.164K}. Due to eROSITA's shallow survey exposures, it detects 3--5 times fewer counts for XINS candidates that are an order of magnitude brighter than those in the 4XMM-DR9 sample. However, for typical eROSITA XINS candidates (with approximately 60 counts in the 0.2–2 keV energy band), our pilot study shows that robust spectral information can still be obtained at these count levels.

Looking ahead, enhanced detection algorithms, alongside multiwavelength and variability characterisation of soft X-ray sources, will continue to aid in identifying the best XINS candidates. The next phase of the project will focus on the scientific analysis of the 4XMM-DR9 XINS sample, aiming to constrain neutron star distances and better understand their distribution in the Galaxy. These insights, in combination with evolutionary models and extinction maps, will contribute significantly to our understanding of neutron star populations beyond the Solar neighbourhood.

\newpage
\section*{Acknowledgments}

We acknowledge the XMM2ATHENA Consortium for providing supporting material on various aspects of the work presented here. The work of A.M.P.~is supported by the Innovation and Development Fund of Science and Technology of the Institute of Geochemistry, Chinese Academy of Sciences, the National Key Research and Development Program of China (Grant No.~2022YFF0503100), the Strategic Priority Research Program of the Chinese Academy of Sciences (Grant No. XDB 41000000), and the Key Research Program of the Chinese Academy of Sciences (Grant NO. KGFZD-145-23-15). I.T.~is supported by Deutsches Zentrum f\"ur Luft- und Raumfahrt (DLR) through grant and 50\,OX\,2301.
The XMM2ATHENA project has received funding from the European Union's Horizon 2020 research and innovation programme under grant agreement no.~101004168.


\bibliography{ref_insX2A}%

\end{document}